\documentclass[nonacm,sigconf,screen]{acmart}

\usepackage{amsmath}
\usepackage{booktabs}

\usepackage{algorithm}
\usepackage{algpseudocode}

\author{Yifan Zhu}
\orcid{0009-0000-6718-7787}
\affiliation{
  \institution{University of Rochester}
  \streetaddress{Rochester, NY}
  \country{USA}
}
\email{yifanzhu@rochester.edu}

\author{Yekai Pan}
\orcid{0009-0001-7639-1488}
\affiliation{
  \institution{University of Rochester}
  \streetaddress{Rochester, NY}
  \country{USA}
}
\email{ypan34@u.rochester.edu}

\author{Chen Ding}
\orcid{0000-0003-4968-6659}
\affiliation{
  \institution{University of Rochester}
  \streetaddress{Rochester, NY}
  \country{USA}
}
\email{cding@cs.rochester.edu}

\algnewcommand\algorithmicparfor{\textbf{parallel for}}
\algdef{S}[FOR]{ParFor}[1]{\algorithmicparfor\ #1\ \algorithmicdo}

\AtBeginDocument{%
  }

\begin{document}

\title{Sawtooth Wavefront Reordering}
\subtitle{Enhanced CuTile FlashAttention on NVIDIA GB10}


\begin{abstract}
High-performance attention kernels are essential for Large Language Models. 
This paper presents analysis of CuTile-based Flash Attention memory behavior and a technique to improve its cache performance.  In particular, 
our analysis on the NVIDIA GB10 (Grace Blackwell) identifies the main cause of L2 cache miss. 
Leveraging this insight, we introduce a new programming technique called Sawtooth Wavefront Reordering that reduces L2 misses. We validate it in both CUDA and CuTile, observing 50\% or greater reduction in L2 misses and up to 60\% increase in throughput on GB10.
\end{abstract}


\maketitle

\section{Introduction}
\label{sec:introduction}

The Attention mechanism is the computational core of transformer-based generative models~\cite{AttentionIsAllYouNeed, GPT3}. To address its quadratic memory complexity, techniques like Flash Attention~\cite{FlashAttention} utilize IO-aware tiling to maximize data reuse in on-chip SRAM. Concurrently, programming models such as NVIDIA's CuTile~\cite{CUDATile} facilitate the development of such tiled algorithms by abstracting complex hardware details.

While these advancements have democratized high-performance kernel development, the abstraction can obscure the nuanced interactions between thread scheduling and the specific cache hierarchy of modern architectures. Prior work has highlighted the importance of hardware thread scheduling and locality awareness in cache management~\cite{10.1109/MICRO.2012.16, 6835937, 10.1145/3037697.3037709}. 


In this work, we improve the performance of CuTile-based Flash Attention. Our approach begins by using hardware counters and performance models to identify the primary cause of non-compulsory cache misses.  
We use raw CUDA implementation with specifically designed Thread Block (CTA) scheduling to isolate the effects of memory access from other compiler-induced effects.

Our analysis reveals that the L1 cache provides negligible benefit for streaming attention patterns, and that L2 cache behavior follows a deterministic model. We identify a correlation between L2 hit rates and the number of active SMs, suggesting intrinsic data reuse among synchronous wavefronts. Based on these findings, we propose a ``Sawtooth'' alternating scan pattern that increases L2 cache data reuse. Finally, we port this optimization to the CuTile environment, demonstrating that the insights gained from low-level analysis translate directly to improvements in the high-level programming model.



\section{Background}
\label{sec:background}

\subsection{GPU Memory Hierarchy}

NVIDIA GPUs are composed of multiple Streaming Multiprocessors (SMs). The CUDA programming model abstracts the parallel execution units as Cooperative Thread Arrays (CTAs), or thread blocks, which are scheduled onto SMs.

The memory hierarchy consists of global memory, L2 cache, and the per-SM L1 memory system. The L1 level is often referred to as unified L1/Texture cache (L1Tex) \cite{PTXISA}. Importantly, this L1 memory can be partitioned into L1 cache and shared memory. 
In high-performance kernels like Fused Multi-Head Attention, developers predominantly rely on \textbf{Shared Memory} rather than the opaque L1 cache. Explicit management allows for deterministic data placement and reuse patterns, essential for the tiling strategies that Flash Attention relies on.




\paragraph{NVIDIA GB10}
For all our experiments, we use NVIDIA's latest GB10 processor, which was announced in January 2025 and became available in October 2025. GB10 is a Grace Blackwell chip fabricated on TSMC's 3nm process node, featuring a unified architecture that combines a Blackwell GPU with 48 streaming multiprocessors and 20 ARM v9.2 CPU cores \cite{skende2025NVIDIA}. The chip has a 256b bus LPDDR5X 128GB unified memory, with $\sim$301GB/s raw bandwidth, and an aggregate bandwidth of $\sim$600GB/s. The graphics processor has an L2 cache size of 24MiB. 

\paragraph{Nsight Compute CLI}
The Nsight Compute CLI (ncu) \cite{NVIDIA_nsight_compute_cli} is a command-line tool 
that provides detailed performance metrics and analysis at the kernel level. The tool offers comprehensive metrics including memory throughput, instruction execution statistics, and occupancy information, making it invaluable for identifying performance bottlenecks and optimizing CUDA kernels. For our experimental purpose, we use the following metrics: \texttt{lts\_\_t\_sectors.sum} and \texttt{lts\_\_t\_sector\_hit\_rate.pct}, corresponding to the total number of sectors requested for any operation and the number of cache hit of from these requests respectively.

\subsection{Flash Attention and Implmementation}
\label{sec:kernels}

Standard scaled dot-product attention computes the output matrix $\mathbf{O}$ as:
\begin{equation}
 \mathbf{O} = \text{softmax}\left(\frac{\mathbf{Q}\mathbf{K}^\top}{\sqrt{d}}\right)\mathbf{V}
\end{equation}
where $\mathbf{Q}, \mathbf{K}, \mathbf{V} \in \mathbb{R}^{N \times d}$ are the query, key, and value matrices. This involves materializing the intermediate attention matrix $\mathbf{S} = \mathbf{Q}\mathbf{K}^\top$ and $\mathbf{P} = \text{softmax}(\mathbf{S})$, which are of size $N \times N$. Since $N$ is typically the sequence length, this results in $O(N^2)$ memory complexity.

FlashAttention \cite{FlashAttention} avoids materializing these large intermediate matrices by fusing the computation. It computes the attention output block-by-block. By keeping small tiles of $\mathbf{Q}, \mathbf{K}, \mathbf{V}$ and intermediate accumulators in the fast on-chip SRAM, it significantly reduces the number of read/write operations to the slow global memory (HBM), practically achieving linear memory complexity with respect to sequence length.

We implement and evaluate two Flash Attention kernel variants utilizing WMMA Tensor Core operations. Our study focuses specifically on L2 cache behavior, utilizing a \textbf{split-Q} dataflow where Query tiles remain resident in shared memory while Key and Value tiles are streamed from global memory.

\paragraph{Tiling Implementation}
\label{sec:kernel-algorithm}

Given query, key, and value matrices $\mathbf{Q}, \mathbf{K}, \mathbf{V} \in \mathbb{R}^{N \times d}$, we employ a tiled approach. Consistent with our implementation code, we enforce \textbf{square tiling} where the block size for the query dimension ($B_r$) equals the block size for the key-value dimension ($B_c$), denoted as $T$.

For each resident query tile $\mathbf{Q}_i \in \mathbb{R}^{T \times d}$, we stream through blocks of keys and values $\{\mathbf{K}_j, \mathbf{V}_j\}_{j=1}^{T_c}$. The logic follows the standard Flash Attention forward pass:

\begin{algorithm}
\caption{Split-Q Fused Multihead Attention with Square Tiling}
\begin{algorithmic}[1]
\State \textbf{Input:} $\mathbf{Q}, \mathbf{K}, \mathbf{V} \in \text{Global Memory}$
\State \textbf{Config:} Determine max square tile size $T \times T$ based on SRAM
    \ParFor{$i = 1$ to $T_r$} \Comment{Grid-stride loop over Q}
    \State Load $\mathbf{Q}_i$ into Shared Memory (Resident)
    \State Initialize $\mathbf{O}_i = \mathbf{0}, \ell_i = \mathbf{0}, m_i = -\infty$
    \For{$j = 1$ to $T_c$} \Comment{Stream K, V tiles}
        \State Load $\mathbf{K}_j, \mathbf{V}_j$ into separate Shared Memory buffers
        \State Compute $\mathbf{S}_{ij} = \mathbf{Q}_i \mathbf{K}_j^T$ (WMMA)
        \State Update Softmax stats $m_i, \ell_i$ (Online Softmax)
        \State Compute $\mathbf{P}_{ij} = \text{softmax}(\mathbf{S}_{ij} \cdot \text{scale}, m_i, \ell_i)$
        \State Compute $\mathbf{O}_{i} \leftarrow \mathbf{O}_{i} + \mathbf{P}_{ij} \mathbf{V}_j$ (WMMA)
    \EndFor
    \State Write $\mathbf{O}_i$ to Global Memory
\EndFor
\end{algorithmic}
\end{algorithm}

\paragraph{CTA Scheduling and Work Distribution}
\label{sec:cta-scheduling}

To manage occupancy and load balancing across the GPU, we employ a \textbf{persistent CTA} (Cooperative Thread Array) pattern with round-robin tile assignment. This gives a deterministic way to study the effect of scheduling on L2 cache.

Instead of launching a thread block for every query tile (which could lead to tail effects where some SMs are idle while others finish the last wave), we launch a fixed number of thread blocks. Each thread block (CTA) executes a grid-stride loop, claiming available Query tiles in a round-robin fashion. This ensures that all SMs remain active until the entire workload is near completion.

\begin{algorithm}
\caption{Persistent CTA Scheduling (Grid-Stride Loop)}
\begin{algorithmic}[1]
\State \textbf{Input:} Total Query Tiles $N_{tiles}$, Grid Size $G$
\State \textbf{Given:} GPU SM Count $N_{SM}$
\State $G \leftarrow \min(N_{tiles}, N_{SM})$ \Comment{One persistent CTA per SM}
\State $k \leftarrow \text{blockIdx.x}$
\While{$k < N_{tiles}$}
    \State Identify ($Batch, Head, TileIndex$) from linear index $k$
    \State Execute \textbf{Algorithm 1} for this Query tile
    \State $k \leftarrow k + G$ \Comment{Stride by total number of CTAs}
\EndWhile
\end{algorithmic}
\end{algorithm}

Alternatively, we can employ an approach where CTAs are launched directly based on the work size. In this scheme, the grid size is set exactly to cover the total number of query tiles (potentially split by batch and heads). This relies on the GPU's hardware scheduler to distribute blocks to SMs.

\begin{algorithm}
\caption{Non-Persistent Scheduling}
\begin{algorithmic}[1]
\State \textbf{Input:} Grid Dim X (Num Q Tiles), Grid Dim Y (Batch * Heads)
\State \textbf{Input:} $\mathbf{Q}, \mathbf{K}, \mathbf{V}$ tensors
\State $q\_tile \leftarrow \text{blockIdx.x}$
\State $batch\_head \leftarrow \text{blockIdx.y}$
\If{$q\_tile \ge num\_q\_tiles$ \textbf{or} $batch\_head \ge batch\_size \times num\_heads$}
    \State \Return
\EndIf
\State Identify $\mathbf{Q}_{q\_tile}$ for current batch/head
\State Load $\mathbf{Q}$ tile into Shared Memory (Resident)
\State ... (Proceed with K, V streaming as in Algorithm 1)
\State Write Output $\mathbf{O}$
\end{algorithmic}
\end{algorithm}

\section{Cache Performance Analysis}

The section presents three studies of the cache behavior of the flash attention kernel.

\subsection{Effect of L1 Caching}

To isolate and model the L2 cache behavior accurately, we must first determine the extent to which the L1 cache filters memory requests. In traditional CPU architectures, the L1 cache acts as a primary filter, absorbing a significant portion of traffic and complicating the prediction of L2 access patterns.

For GPU attention kernels, however, the heavy reliance on shared memory suggests a different interaction. Although $Q$ tiles are resident in shared memory, they are streamed from global memory during the outer loop. Similarly, $K$ and $V$ tiles exhibit a streaming pattern with even shorter tenancy in shared memory. We hypothesize that for these streaming access patterns, the L1 cache essentially functions as a pass-through buffer. Verifying this hypothesis is a necessary first step, as it allows us to simplify our L2 modeling by treating L2 access counts as a direct function of global memory requests.

\begin{figure}
    \centering
    \includegraphics[width=0.75\linewidth]{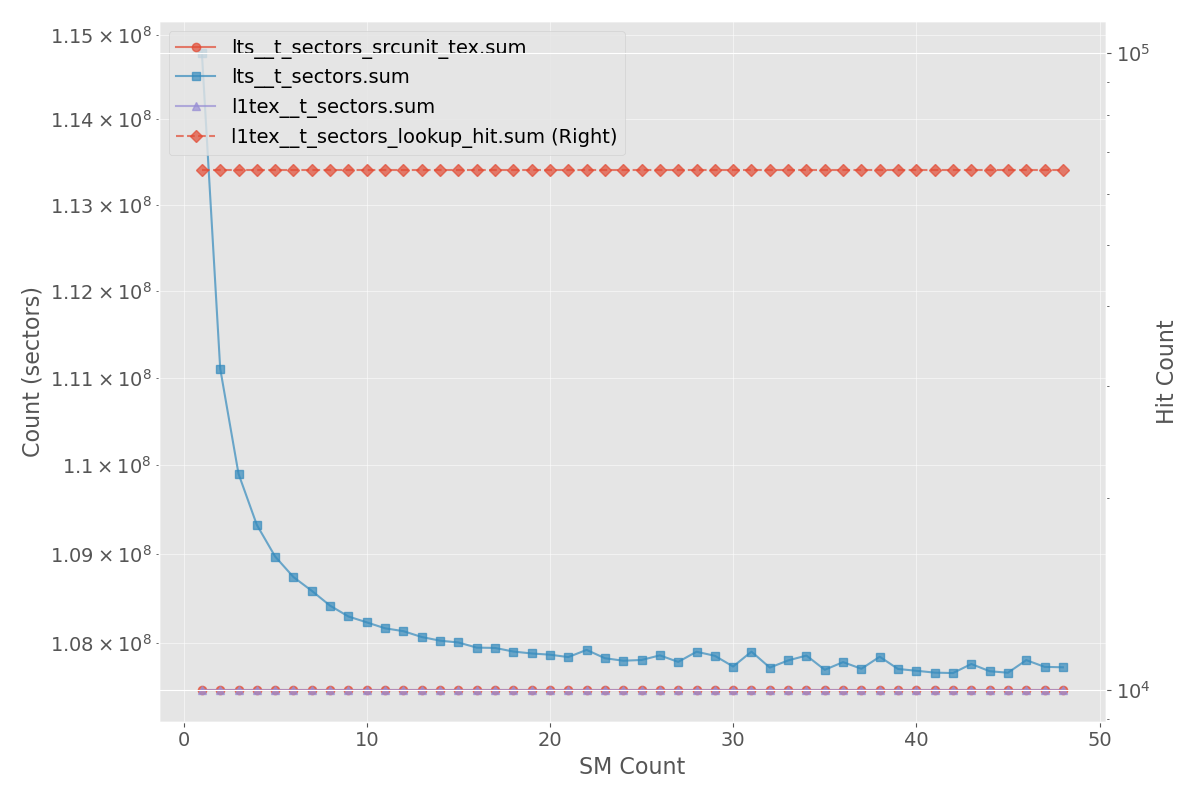}
    \caption{L1/L2 Metrics for Sequence Length 32K. Parameters: Batch=1, Heads=1, Head Dim=64, Tile Size=80x80.}
    \label{fig:l1-l2-32k}
\end{figure}

\begin{figure}
    \centering
    \includegraphics[width=0.75\linewidth]{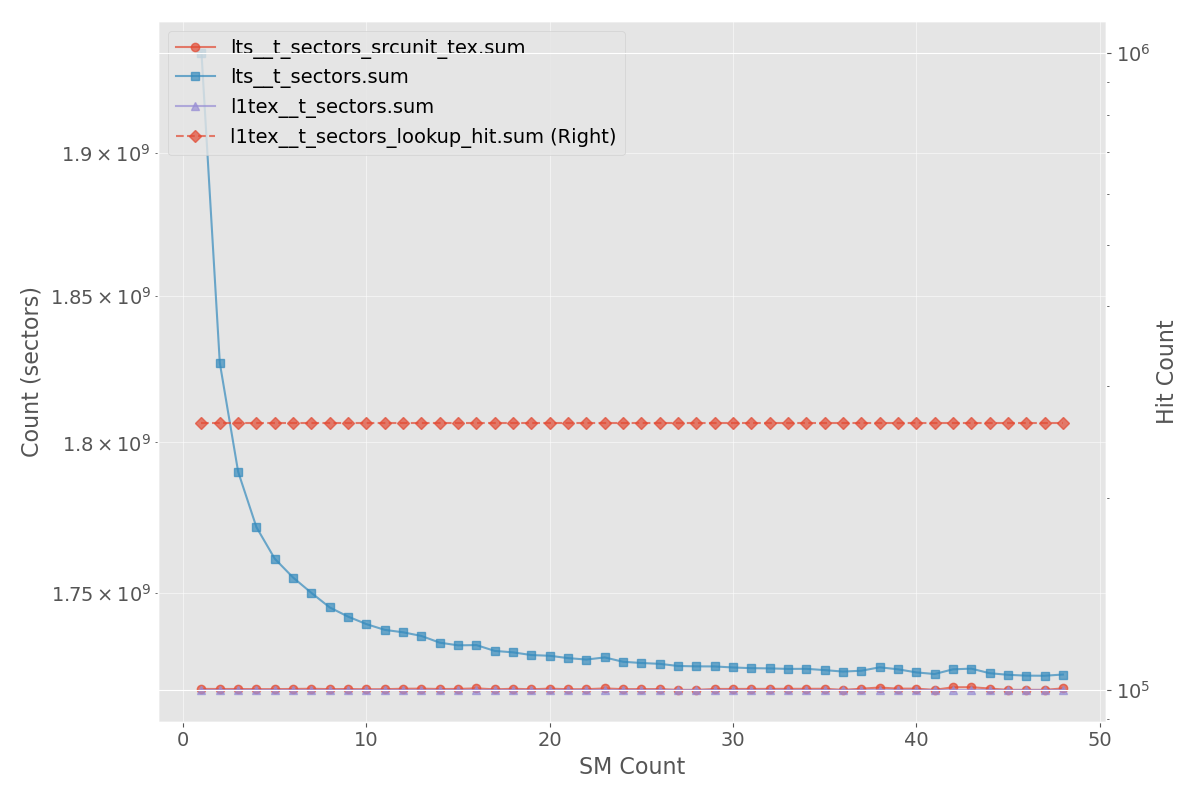}
    \caption{L1/L2 Metrics for Sequence Length 128K. Parameters: Batch=1, Heads=1, Head Dim=64, Tile Size=80x80.}
    \label{fig:l1-l2-128k}
\end{figure}

To zoom into the specific quantitative behavior, Table \ref{tab:l1-l2-counters} lists the exact counter values for the fully saturated case (SM=48).

\begin{table}[h]
\centering
\caption{L1/L2 Cache Counters for SM=48}
\label{tab:l1-l2-counters}
\begin{tabular}{lrr}
\toprule
\textbf{Metric} & \textbf{32K Seq Len} & \textbf{128K Seq Len} \\
\midrule
L2 Sectors (Total) & 107,729,467 & 1,723,556,561 \\
L2 Sectors (from Tex) & 107,478,656 & 1,719,093,980 \\
L1 Sectors (Total) & 107,478,656 & 1,718,615,808 \\
L1 Hit Count & 65,440 & 262,080 \\
\bottomrule
\end{tabular}
\end{table}

We also collected data for the non-persistent CTA scheduling case (classic grid launch) to determine if thread block scheduling impacts these cache behaviors. The results are shown in Table \ref{tab:l1-l2-counters-classic}.

\begin{table}[h]
\centering
\caption{L1/L2 Cache Counters for Non-Persistent CTA (SM=48)}
\label{tab:l1-l2-counters-classic}
\begin{tabular}{lrr}
\toprule
\textbf{Metric} & \textbf{32K Seq Len} & \textbf{128K Seq Len} \\
\midrule
L2 Sectors (Total) & 107,991,698 & 1,723,401,754 \\
L2 Sectors (from Tex) & 107,741,184 & 1,719,664,640 \\
L1 Sectors (Total) & 107,741,184 & 1,719,664,640 \\
L1 Hit Count & 65,536 & 262,144 \\
\bottomrule
\end{tabular}
\end{table}

The data indicates that the non-persistent case exhibits nearly identical L1/L2 behavior to the persistent case. The L1 hit count remains negligible, and the L2 usage is overwhelmingly driven by traffic from the L1 texture path (\textsc{L1Tex}). This suggests that for this streaming workload, provided the GPU is fully saturated (SM=48), the specific software scheduling mechanism (persistent vs. non-persistent) has minimal impact on the fundamental memory access patterns and cache efficiency.  Here are the key observations.

\begin{enumerate}
    \item \textsc{L1Tex} hit count is tiny (hence visualized on a separate axis). This confirms that streaming data effectively bypasses the benefits of L1 caching.
    \item Although L2 total cache sector access has some overhead, L1 miss (traffic from L1 to L2) is the main contributor to L2 traffic.
    \item This behavior is observed in both small (32K) and large (128K) sequence lengths.  It is consistent across different SM counts, though L2 overhead varies slightly.
\end{enumerate}

\subsection{Modeling L2 Sector Access}

We establish a formula to compute the L2 sector access. For now, we only focus on single-batch single-head case, as these two parameters are two linear factors that scale the problem. We use the following variables:

\begin{itemize}
    \item $S$: Sequence Length
    \item $C$: Sector Size
    \item $E$: Element Size
    \item $T$: Tile Size
    \item $D$: Head Dimension
    \item $M$: Number of Sectors
\end{itemize}

Each tile consists of $TD$ elements, hence $\frac{TDE}{C}$ sectors. For $Q$ and $O$, each tile is only accessed once, resulting in $2\frac{TDE}{C}  \left\lfloor\frac{S}{T}\right\rfloor$ sectors. Without causal masking, $K$ and $V$ tiles are accessed once per $Q$ tile, resulting in $2\frac{TDE}{C}  \left\lfloor\frac{S}{T}\right\rfloor^2$ sectors. The total access count is thus the sum of $Q, K, V, O$ accesses plus the trailing incomplete tile accesses.

Ignoring the trailing effect and utilizing direct division to approximate, we have $M = 2\left(\frac{SDE}{C} + \frac{S^2DE}{TC}\right)$.

On our experimental device (DGX Spark) with float16, we have $C = 32$, $E = 2$ and we fix $D = 64$ in this study, hence, we get:
$$
M \approx 8S\left(1 + \frac{S}{T}\right)
$$

With causal masking, the access count for $K$ and $V$ changes from $\left(\frac{S}{T}\right)^2$ to $\frac{S(S-1)}{2T}$, hence we have:
$$
M \approx 8S\left(\frac{S}{2T} + \frac{1}{2}\right)
$$

\begin{figure}
    \centering
    \includegraphics[width=0.75\linewidth]{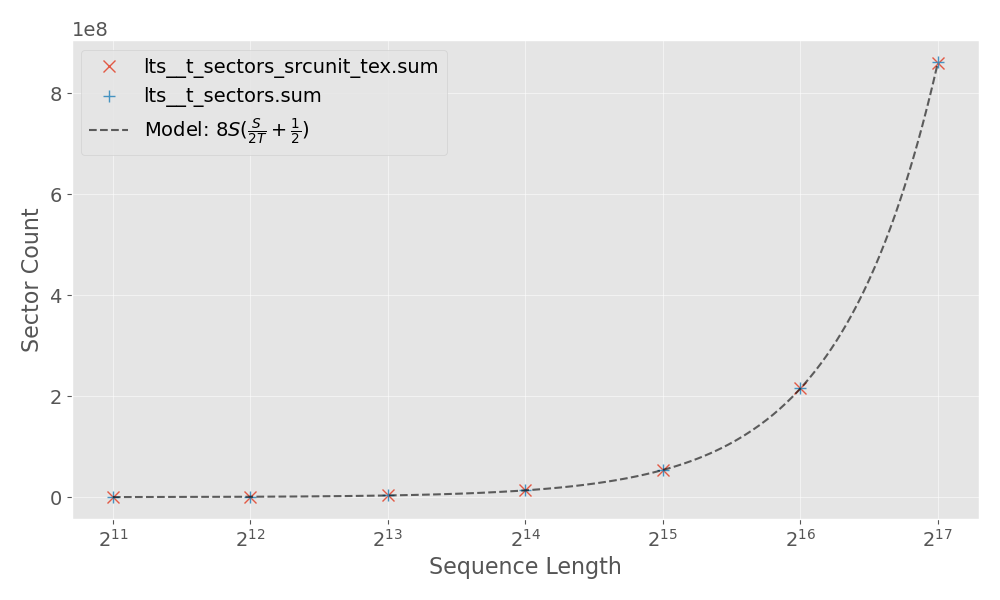}
    \caption{L2 Sector Access vs Sequence Length (Causal Masking, $T=80$).}
    \label{fig:study2-causal}
    \label{fig:study2-plots}
\end{figure}

We validate these models against experimental data in Figure \ref{fig:study2-plots}. The results confirm that our formulas accurately predict the L2 sector access count for both scenarios. In this specific experiment, the tile size was $T=80$.

\begin{figure}
    \centering
    \includegraphics[width=0.75\linewidth]{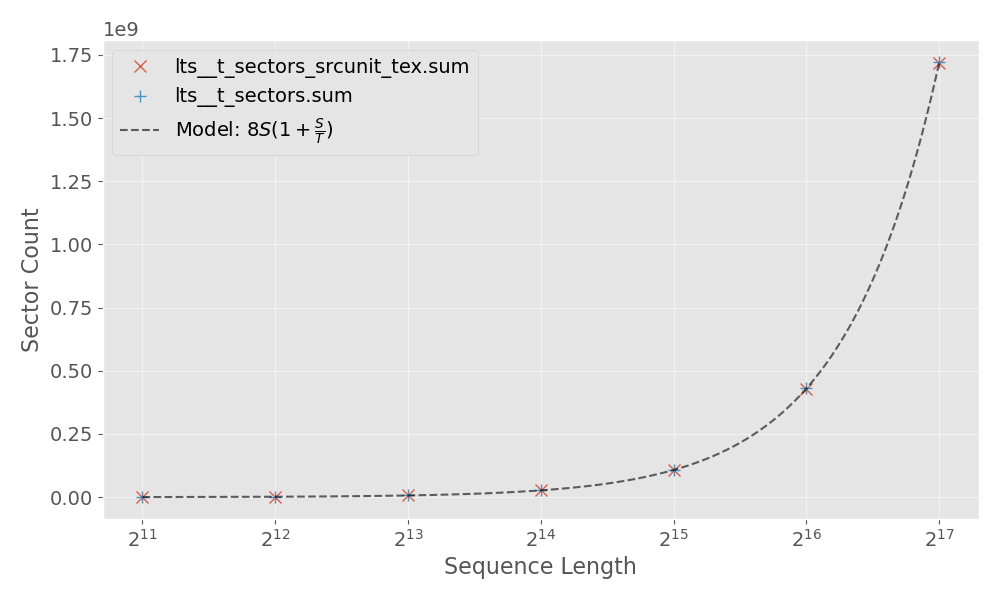}
    \caption{L2 Sector Access vs Sequence Length (Non-Causal Masking, $T=80$).}
    \label{fig:study2-regular}
\end{figure}

To further quantify the accuracy of our model, we analyzed the Mean Absolute Percentage Error (MAPE) between the experimental L2 sector counts and our theoretical predictions for the saturated SM=48 case. As shown in Table \ref{tab:mape-results}, the experimental data fits the model extremely well, with less than 1\% error for the non-causal case and minimal deviation for the causal case. Both the total L2 sectors and the subset originating from the texture unit (L1Tex misses) fall almost exactly on the predicted curve.

\begin{table}[h]
\centering
\caption{MAPE of Theoretical Model vs Experimental Data (SM=48)}
\label{tab:mape-results}
\begin{tabular}{lrr}
\toprule
\textbf{Metric} & \textbf{Non-Causal(\%)} & \textbf{Causal (\%)} \\
\midrule
L2 Sectors (Total) & 0.4527\% & 2.4941\% \\
L2 Sectors (from Tex) & 0.5389\% & 1.1286\% \\
\bottomrule
\end{tabular}
\end{table}

\subsection{L2 Non-Compulsory Miss Threshold}

The L2 capacity is quite limited compared to the requirements of large sequence lengths. Hence, it is important to study the onset of L2 non-compulsory misses.

If the L2 cache can hold all the data, we should see the L2 sector miss roughly equal to the cold miss count. In this scenario, since $Q, K, V, O$ are the major four tensors, the cold miss should approximately equal to $4\frac{SDE}{C}$, which is $16S$ with our configuration.

\begin{figure}
    \centering
    \includegraphics[width=0.75\linewidth]{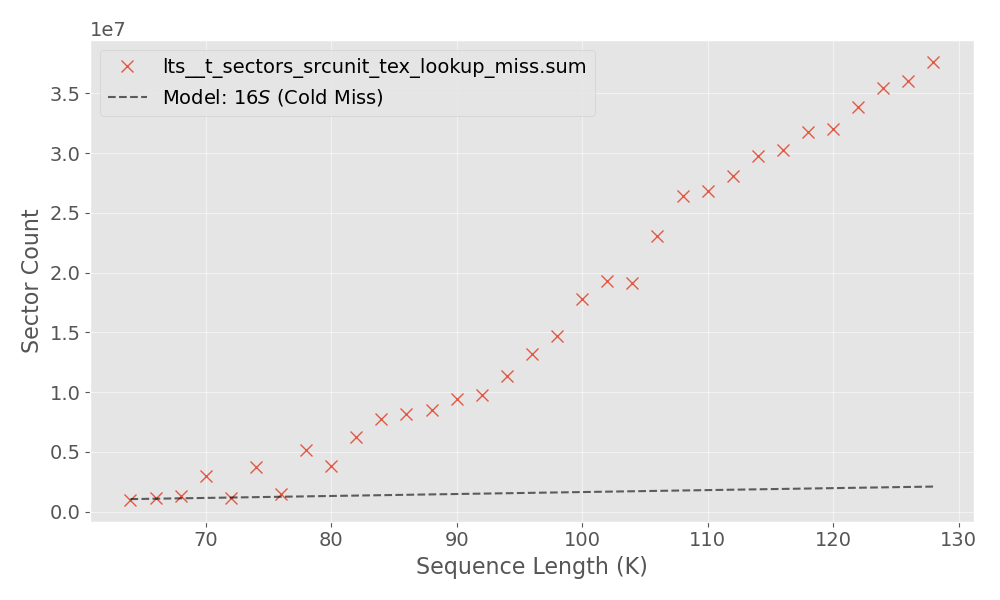}
    \caption{L2 Miss Count vs Sequence Length (SM=48). The dashed line represents the theoretical cold miss count ($16S$).}
    \label{fig:study3-miss}
\end{figure}

Given that the KV matrix constitutes a major portion of the access volume, we expect L2 misses to diverge from cold misses only when the KV size approached the L2 cache size. Experimental results in Figure~\ref{fig:study3-miss} indicate that the divergence point occurs at a sequence length ($S$) of approximately 80K, corresponding to a KV size of 20 MiB, which is expected as the cache size is 24MiB.

\subsection{The Cause of L2 Non-Compulsory Miss}

\begin{figure}
    \centering
    \includegraphics[width=0.75\linewidth]{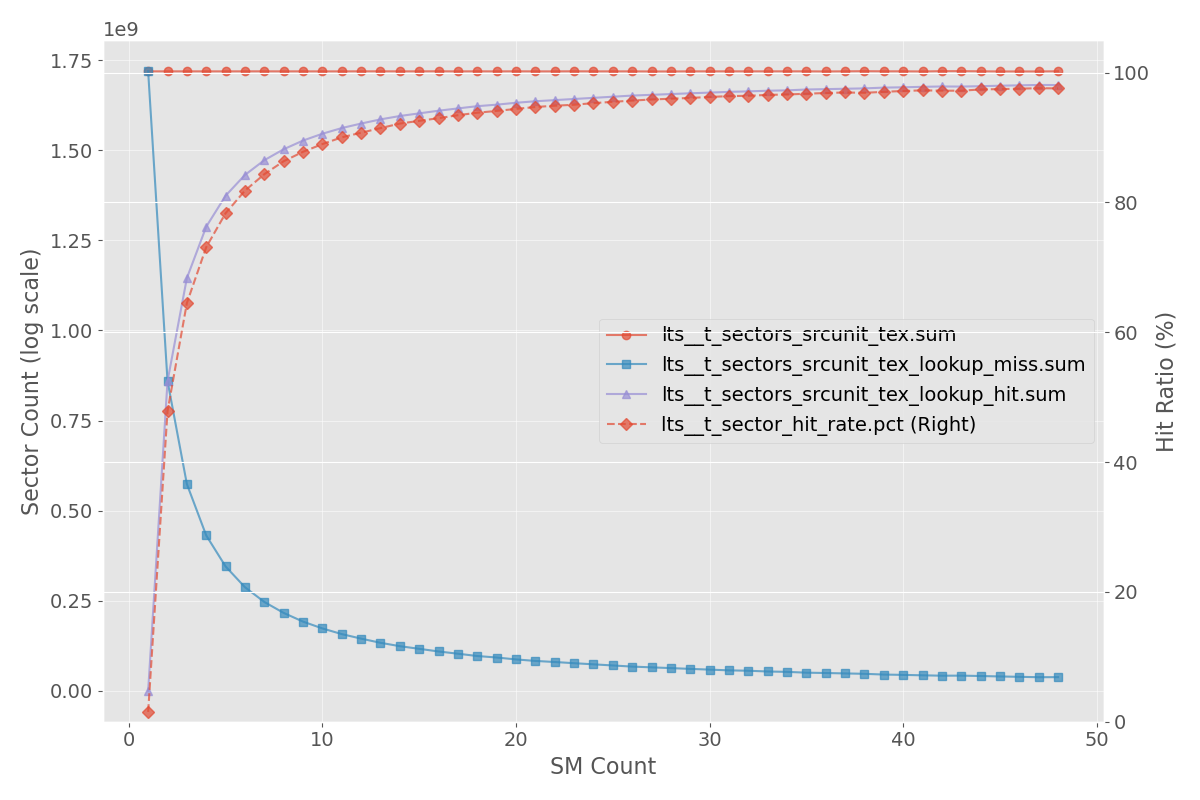}
    \caption{L2 Cache Miss Count and Hit Rate vs. Number of Active SMs. The scale factor for the hit rate is observed to be approximately $1 - \frac{1}{N_{SM}}$.}
    \label{fig:study4-miss}
\end{figure}

In this study, we examine the relationship between the number of active Streaming Multiprocessors (SMs) and L2 cache behavior, with a particular focus on \emph{non-compulsory misses} (i.e., misses beyond cold misses). Figure~\ref{fig:study4-miss} presents the L2 miss count and L2 hit rate as functions of the number of active SMs.

As the number of SMs increases, a clear trend appears in the L2 hit rate. The experimental results suggest that the hit rate scales with the number of SMs (\(N_{SM}\)) approximately according to the factor \(1 - \frac{1}{N_{SM}}\). This scaling implies that activating more SMs increases the likelihood that a requested cache line is already present in the shared L2 cache. The final hit ratio approaches \(1 - \frac{1}{48} \approx 98\%\) for this architecture.

This behavior is an indicator of \emph{wavefront-like reuse among CTAs}. On the GB10 architecture under this workload, CTAs appear to progress in a largely synchronized manner. As a result, when the first CTAs populates the L2 cache during its global-to-shared memory loads, subsequent CTAs can reuse the same cache lines, leading to L2 hit events instead of misses.

\section{Sawtooth Wavefront Reordering}

The previous studies have established that standard cyclic access patterns lead to significant L2 non-compulsory misses once the KV data size exceeds the L2 capacity. This inefficiency arises because the reuse distance--the volume of data accessed between two reuses of the same cache line--exceeds the cache size.  Reuse distance is originally called the LRU stack distance~\cite{Mattson+:IBM70}.

To reduce the reuse distance, we change the code to iterate the KV data in a different order, from cyclic to sawtooth.  In particular, we alternate the scanning direction of the inner loop (loading KV blocks from 0 to $N$ in even iterations and $N$ to 0 in odd iterations).  Unlike the cyclic order where all reuse distances equal to the data size, the sawtooth order reduces the reuse distance for most data accesses to be less than the data size.  We call this technique \emph{Sawtooth Wavefront Reordering}.

In this section, we first implement and evaluate this optimization in our controlled CUDA environment to quantify the raw theoretical gain. Subsequently, we port this logic to the CuTile implementations to validate that these low-level insights translate effectively to abstracted programming models.

\subsection{Implementation}
The implementation distributes the workload by assigning sequences of Q tiles to each SM. We present the logic for the alternating scan pattern in Algorithm \ref{alg:sawtooth}.

\begin{algorithm}[H]
\caption{Sawtooth KV Access Pattern}
\label{alg:sawtooth}
\begin{algorithmic}[1]
\State \textbf{Input:} Sequence of Query tiles $Q_{seq}$ assigned to SM
\State \textbf{Input:} Total KV tiles $N_{KV}$
\State $i_{local} \gets 0$
\For{each query tile $q$ in $Q_{seq}$}
    \State \Comment{Determine scan direction based on local iteration parity}
    \If{$i_{local} \pmod 2 = 0$}
        \State $start \gets 0$, $end \gets N_{KV}$, $step \gets 1$ \Comment{Forward}
    \Else
        \State $start \gets N_{KV}-1$, $end \gets -1$, $step \gets -1$ \Comment{Backward}
    \EndIf
    \State
    \For{$j \gets start$ \textbf{to} $end$ \textbf{step} $step$}
        \State Load KV tile $K_j, V_j$
        \State Compute Attention($q, K_j, V_j$)
    \EndFor
    \State $i_{local} \gets i_{local} + 1$
\EndFor
\end{algorithmic}
\end{algorithm}

\subsection{CUDA Results}

\begin{figure}
    \centering
    \includegraphics[width=0.75\linewidth]{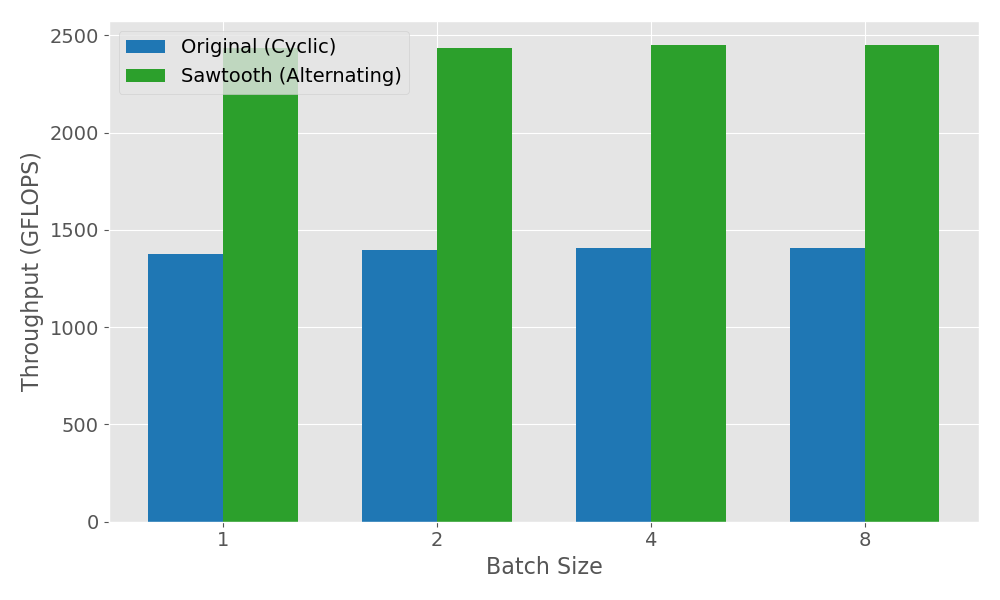}
    \caption{Kernel Throughput: Original (Cyclic) vs. Sawtooth.  
    }
    \label{fig:study5-throughput}
\end{figure}

\begin{figure}
    \centering
    \includegraphics[width=0.75\linewidth]{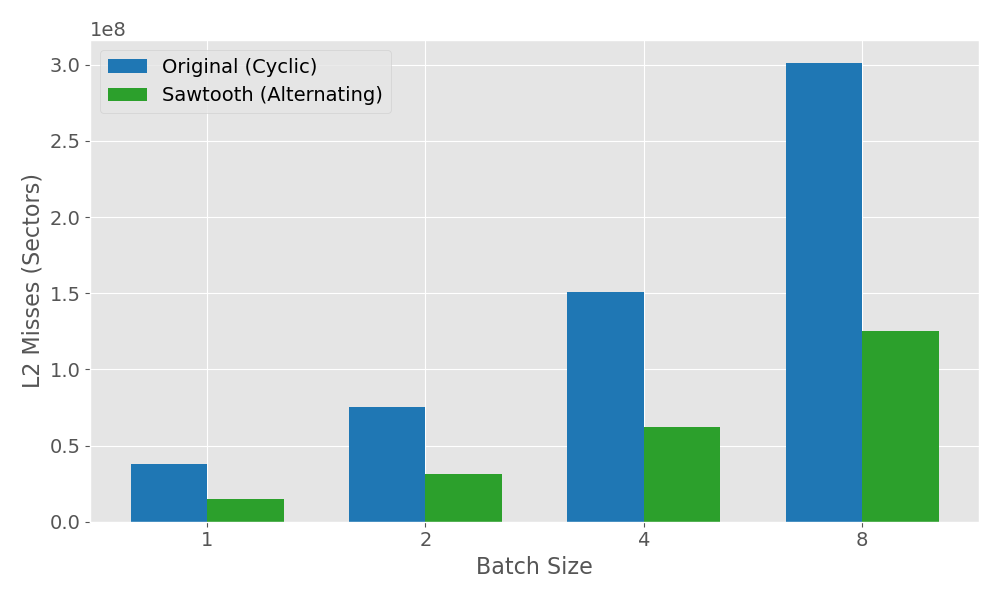}
    \caption{L2 Cache Misses: Original (Cyclic) vs. Sawtooth.  
    }
    \label{fig:study5-miss}
\end{figure}

We evaluated the performance of Sawtooth Wavefront Reordering against the baseline (cyclic) approach across varying batch sizes ($B=\{1, 2, 4, 8\}$). As shown in Figure \ref{fig:study5-throughput} and \ref{fig:study5-miss}, the proposed method achieves significant reductions in L2 cache misses and corresponding improvements in throughput.

Sawtooth Wavefront Reordering reduces the L2 non-compulsory misses by approximately 50\% across all tested configurations. This reduction in memory traffic translates directly into performance gains, with throughput increasing significantly (e.g., from approx. 1.3 TFLOPS to 2.4 TFLOPS). This confirms that properly managing the reuse distance of large streaming buffers (like KV cache) can effectively mitigate non-compulsory misses even when the working set exceeds the cache size.

\subsection{Validation on CuTile}
We further validated the L2 non-compulsory miss reduction optimization by porting the Sawtooth optimization to \texttt{CuTile}, a Python-based tile-centric programming environment. NVIDIA CUDA Tile is a tile-based GPU programming model that targets portability for NVIDIA Tensor Cores. CUDA Tile achieves peak GPU performance with a programming model that simplifies the creation of optimized, tile-based kernels across NVIDIA platforms \cite{CUDATile}. We evaluated two implementation variants:

\begin{itemize}
    \item \textbf{Fully Static}: A direct port of the persistent CTA logic where the entire schedule is statically determined.
    \item \textbf{Tile-based}: A variant that leverages the tile-based programming model. Unlike the static schedule, this variant locally advances the sequence loop by a step of 2 and alternates the order accordingly to achieve the sawtooth pattern.
\end{itemize}

Both programs are implemented with a tile size of $64\times64$  and tested with a batch size of $8$, a sequence length of $128\times1024$, and a head dimension of $64$.

\subsubsection{Results}
We profiled both the standard cyclic pattern (Static, Tile) and sawtooth reordered (Static Alt, Tile Alt) on the same hardware configuration. The results, summarized in Figure \ref{fig:cutile-results}, align closely with our CUDA findings. Sawtooth Wavefront Reordering consistently reduces the total L2 miss count from approximately 370 million to 120 million sectors—a reduction of roughly 67\%. This efficiency gain translates into a measurable performance improvement, with performance increasing from $\sim$61 TFLOPS to $\sim$69 TFLOPS ($\sim$13\%). In the causal variant, the sawtooth pattern is also effective as the inner KV traversal loop still covers mostly the same data for each schedule round of all CTAs. Our experiment shows an increase from $\sim41$ TFLOPS to $\sim$66 TFLOPS ($\sim$60\%).

\begin{figure}
    \centering
    \includegraphics[width=0.75\linewidth]{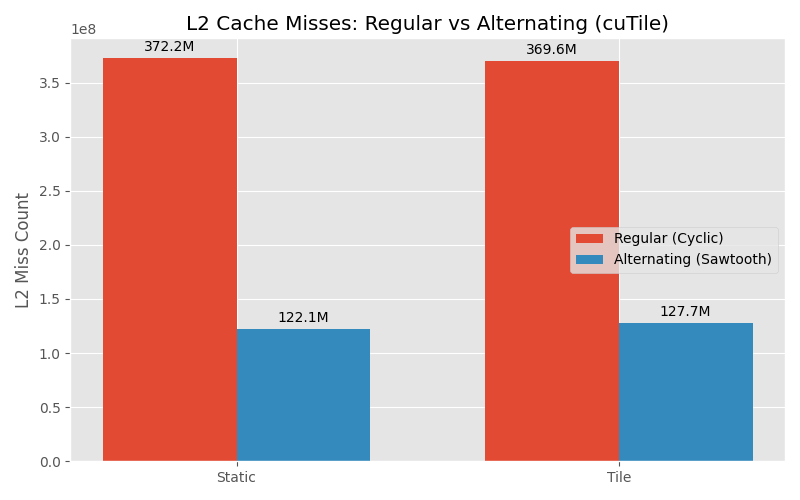}
    \caption{L2 Miss Count Comparison on CuTile without Causal Masking (Regular vs. Sawtooth). The optimization significantly reduces L2 miss traffic for both implementation variants.}
    \label{fig:cutile-misses}
\end{figure}

\begin{figure}
    \centering
    \includegraphics[width=0.75\linewidth]{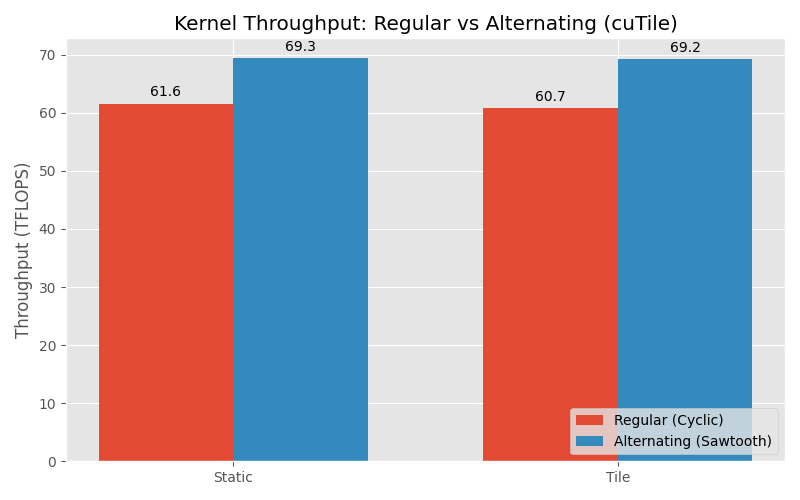}
    \caption{Throughput Comparison on CuTile without Causal Masking (Regular vs. Sawtooth). Reduction in L2 traffic yields $\sim$13\% higher performance.}
    \label{fig:cutile-throughput}
    \label{fig:cutile-results}
\end{figure}

\begin{figure}
    \centering
    \includegraphics[width=0.75\linewidth]{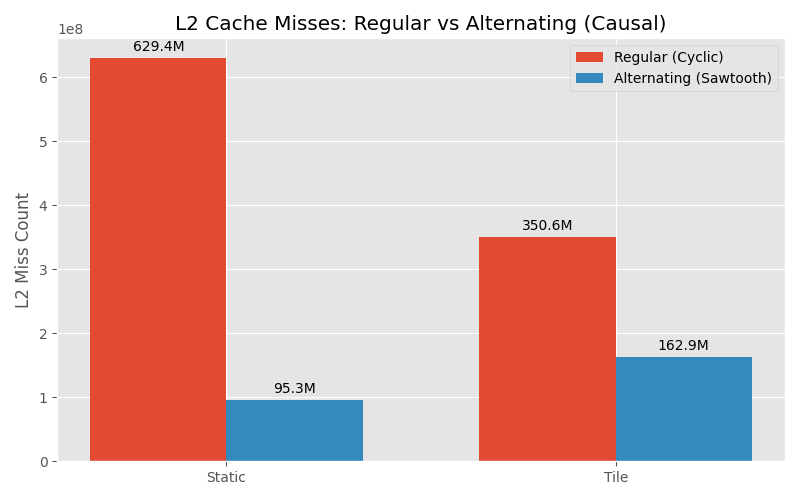}
    \caption{L2 Miss Count Comparison on CuTile with Causal Masking (Regular vs. Sawtooth). The optimization significantly reduces L2 miss traffic for both implementation variants.}
    \label{fig:cutile-misses}
\end{figure}

\begin{figure}
    \centering
    \includegraphics[width=0.75\linewidth]{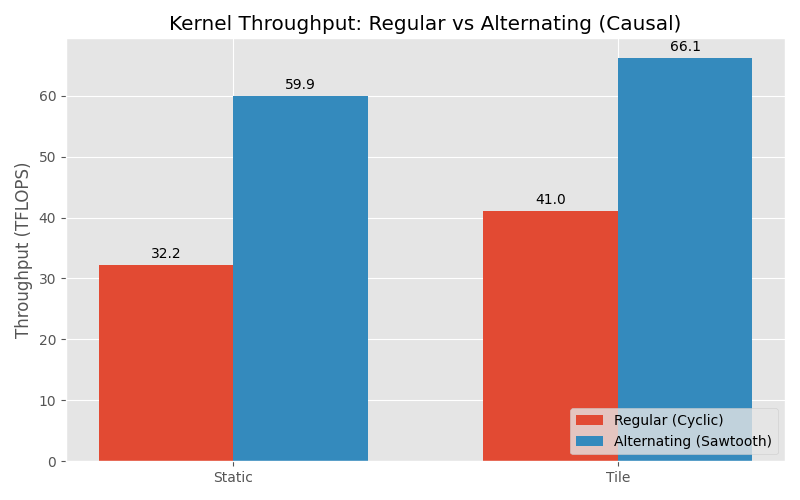}
    \caption{Throughput Comparison on CuTile with Causal Masking (Regular vs. Sawtooth). Reduction in L2 traffic yields $\sim$60\% higher performance.}
    \label{fig:cutile-throughput}
    \label{fig:cutile-results}
\end{figure}

\subsubsection{Limitations}
Although the sawtooth order proves effective, we identified a limitation regarding tile size. The optimization works for regular patterns where the selected tile size is smaller than the shared memory capacity. However, for large tile-based programming models—specifically when testing with a tile size of 128, the CuTile compiler may split large tiles that do not fit in \textsc{L1Tex} into smaller tiles, which alters the access pattern. We leave this for future work.

\section{Related Work}

Cyclic and sawtooth are simplest traversal patterns studied in early work about the working-set model and management~\citep{DenningK:SOSP75}.  The sawtooth pattern has been used as a micro-benchmark to analyze the replacement algorithm on AMD multicore processors built with chipsets and large LLCs~\cite{Sun+:MEMSYS24}.  A recent theory of symmetric locality formalized the set of all retraversal orders~\citep{Escolona+:MEMO24}. 

Sawtooth ordering is a special case of "last-free allocation" used in memory allocators to reuse the most recently freed block for the next request of a compatible size, often implemented by a LIFO free list.  In fact, the call stack is an example of such memory allocation to maximize cache reuse.  As a technique of locality optimization, sawtooth ordering is machine independent, unlike loop tiling which targets a specific cache.  


\section{Summary}
\label{sec:conclusion}

We have presented cache performance analysis and optimization for Flash Attention on GB10.  We have identified the main cause of L2 cache miss and developed Sawtooth Wavefront Reordering to increase data reuse in L2.  Evaluation shows that when applied on the CUDA version of the code, optimization reduces L2 misses by 50\% and increases throughput from 1.3 TFLOPS to 2.4 TFLOPS. In the  CuTile version, it reduces the miss count by 67\% and increases throughput from 61 TFLOPS to 69 TFLOPS (non-causal), 41  TFLOPS to 66 TFLOPS (causal), an increase of 13\% (60\%).

\begin{acks}
We thank Yanghui Wu, Jiaqi Nie, and Yiling Zou for their related work on GPU performance modeling and locality analysis, which has helped inform and motivate our study.
\end{acks}




\bibliographystyle{ACM-Reference-Format}
\bibliography{references,all}

\end{document}